\documentclass[12pt,preprint,3p,onecolumn,sort&compress]{elsarticle}

\DeclareMathAlphabet{\mathpzc}{OT1}{pzc}{m}{it}
\newcommand{\meson}[0]{\mathpzc{m}}
\newcommand{\br}[1]{\;\mathrm{BR}\left(#1\right)}
\newcommand{\unit}[1]{\;\mathrm{#1}}
\renewcommand{\epsilon}{\varepsilon}

\usepackage{hyperref}
\usepackage{physics}
\usepackage{amssymb}
\usepackage{xcolor}

\hypersetup{
  bookmarksopen=true,     
  colorlinks=true,        
  linkcolor=darkBlue,     
  citecolor=darkBlue,      
  filecolor=darkBlue,      
  urlcolor=darkBlue        
}

\journal{Astroparticle Physics}
\begin{document}

\begin{frontmatter}
\title{Meson production in air showers and\\
	the search for light exotic particles}

\author{M.~Kachelrie{\ss} and J.~Tjemsland}

\address{Institutt for fysikk, NTNU, Trondheim, Norway}

\begin{abstract}
Decays of mesons produced in cosmic ray induced air showers in Earth's
atmosphere can lead to a flux of light exotic particles which can be
detected in underground experiments. We evaluate the energy spectra of the 
light neutral mesons $\pi^0$, $\eta$, $\rho^0$, $\omega$, $\phi$ and $J/\psi$
produced in interactions of cosmic ray protons and helium nuclei with air 
using QCD inspired event generators.
Summing up the mesons produced in the individual hadronic interactions
of air showers, we obtain the resulting fluxes of undecayed mesons.
As an application, we re-consider the case of millicharged particles created
in the electromagnetic decay channels of neutral mesons.
\end{abstract}

\end{frontmatter}

\section{Introduction}

While overwhelming evidence has been accumulated for the existence of dark
matter (DM) from astrophysical and cosmological observations, the experimental
searches for such particles in direct detection experiments have
not been successful yet. Combined with the null results in searches for new
physics at the LHC, this indicates that new particles with masses below the
TeV~scale are only weakly coupled to the standard model. The prime candidate
for such a DM particle, a thermal relic with mass around the weak scale, has
been constrained severely and is
on the eve of being excluded: For instance, the upper
limit on the annihilation cross section obtained by the Fermi-LAT
collaboration using dwarf galaxies excludes thermal relics with masses
below $\sim 100$\,GeV~\cite{Ackermann:2015zua}, while model dependent
limits from antiproton data are typically even more
stringent~\cite{Fornengo:2013osa,Kachelriess:2020uoh}.
Therefore, both model building and experimental searches have expanded their
phenomenological scope considerably the last decade,
investigating e.g.\ light DM particles with masses in the sub-GeV range.

Traditionally, this mass range has been considered to be inaccessible to
direct detection experiments, since the recoil energy of a DM
particle with typical Galactic velocities, $v\sim 10^{-3}c$, is below the
threshold energy of such experiments. However,
Refs.~\cite{Bringmann:2018cvk,Ema:2018bih} recently pointed out
that cosmic rays (CRs) colliding with DM can up-scatter them, leading to 
a significantly increased DM
flux above the threshold energy of direct detection experiments.
Another generic source of light DM particles are CR interactions in
the atmosphere of the Earth~\cite{Kusenko:2004qc,Yin:2009yt,Hu:2016xas,%
Arguelles:2019ziu,Coloma:2019htx,Alvey:2019zaa}. If mesons produced in these
interactions decay partially into DM, an energetic DM flux that
can be detected in underground experiments results. While the up-scattering
mechanism relies on a sufficiently large abundance of the DM particle
considered, CR interactions in the atmosphere depend only on the well-known
flux of incident cosmic rays.
This mechanism can moreover produce other long-lived exotic particles,
thereby extending the reach of searches for new physics.

In this work, we re-evaluate the atmospheric fluxes of undecayed $\pi^0$,
$\eta$, $\rho$, $\omega$, $\phi$ and $J/\psi$ mesons, which we denote
collectively by $\meson$.
In a previous study by Plestid {\it et al.}~\cite{Plestid:2020kdm},
these fluxes were computed using parametrisations
for the relevant production cross sections 
in $pp$ collisions. Here, we improve upon this in several aspects:
First, we use  QCD inspired event generators to model the
particle production in single hadronic interactions.
This allows us to account for the contribution of
helium in the CR primary flux as well as for the effect of air
as target nuclei. Comparing the results of different event generators
we obtain an estimate for the uncertainties of their predictions.
Moreover, we model the complete hadronic air shower by
considering interactions of secondaries such as  $\pi^\pm p\to \meson X$ and
$K^\pm p\to \meson X$.  Our main result is thus an
improved description of the atmospheric flux of undecayed mesons
produced in air showers.
Our tabulated results can be used to evaluate the flux of exotic particles
produced by atmospheric meson decays within generic extensions of the standard
model\footnote{%
  Tables for the integrated meson fluxes are attached to the arxiv
  submission.
}.
Possible applications include, for instance,
the decay of $\pi^0$ and $\eta$ mesons into a pair of DM
particles through a bosonic mediator \cite{Alvey:2019zaa},
and the case of millicharged DM that couples to the Standard Model (SM)
via a photon~\cite{Ignatiev:1978xj}.
As an illustration for the application of our atmospheric flux of undecayed
mesons, we consider the production of
a generic millicharged particle
(mCP) and compare our results with those of Ref. \cite{Plestid:2020kdm}. 
Such particles arise naturally through, e.g., the kinetic mixing between
the SM photon and a dark photon \cite{Okun:1983vw,Georgi:1983sy,%
Holdom:1985ag,Dobroliubov:1989mr,Fayet:1990wx}. 
The possible mass-to-charge ratio $m/\epsilon e$ of models in which the DM
is charged are already strongly constrained by astrophysical processes as
well as ground based experiments, see e.g.\
Refs.~\cite{Dunsky:2018mqs,McDermott:2010pa,Kachelriess:2020ams}. 
However, these limits can be avoided, if the charged component is unstable
on cosmological time scales or constitutes only a small part of the total
DM abundance. Therefore, DM theories with a sub-dominant
charged component in a hidden sector have attracted attention,
for recent reviews see Refs.~\cite{Fabbrichesi:2020wbt,Filippi:2020kii}.
An additional motivation for such models is the EDGES anomaly which can
be explained in a small window in parameter space close to the limits
from direct detection experiments~\cite{Bowman:2018yin,Kovetz:2018zan}. 

This paper is structured as follows. In section~\ref{sec:cascade}, we first
compare the meson production cross sections calculated using various QCD
inspired event generators to experimental data, and compute next
the atmospheric flux of undecayed neutral mesons.
As an example for the applicability of the
tabulated fluxes, we re-evaluate the flux of mCPs from atmospheric meson
decays in section~\ref{sec:application}. Finally, a summary is given
in section~\ref{sec:summary}.

\section{Meson production in air showers}
\label{sec:cascade}

High energy cosmic rays entering the atmosphere interact with air nuclei. 
The produced long-lived hadrons will in turn interact with other air nuclei,
thus creating a so-called hadronic air shower. The short-lived particles, on
the other hand, may decay. About 1/3 of the energy is transferred in each
generation of the air shower  into the electromagnetic component, mainly via
the decay of short-lived mesons. Thus, the decay of mesons in a hadronic
air shower may be a promising detection channel for exotic particles that
interact with the SM via a photon, such as mCPs.
To describe the hadronic interactions, we utilize the 
QCD inspired event generators
DMPJET~III~19.1~\cite{Engel:1994vs,Roesler:2000he,Fedynitch:2015kcn},
Pythia~8.303~\cite{Sjostrand:2014zea},
QGSJET~II-04~\cite{Ostapchenko:2010vb,Ostapchenko:2013pia},
Sibyll~2.3d~\cite{Riehn:2017mfm}, and
UrQMD~3.4~\cite{Bass:1998ca}. 
The focus will be on DPMJET and Sibyll which are event generators widely
used in the field of CR physics.
An exception is the production of
$J/\psi$ mesons, where we rely on the event generator Pythia which is focused
on accelerator physics.

\subsection{Production cross sections}
\label{app:cross_section}

Parametrisations of hadronic interactions relying on empirical
scaling laws are often used as an efficient tool to reproduce inclusive
quantities like total cross sections. The use of such
parametrisations
becomes, however, dangerous when they are extrapolated outside the kinematical
range of the data they are based on. Moreover, such parametrisations are
generally not available for the cases where nuclei are employed as
CR primaries or targets.
In this work, we use therefore  Monte Carlo event generators in the description
of hadronic interactions to compute the atmospheric meson fluxes. While this
approach avoids the disadvantages of
parametrisations,
it has also its own
drawbacks: In particular, QCD inspired event generators cannot be used below
a minimal energy, which is typically in the range of 5--10\,GeV/n of
the projectile in the lab frame. While this implies that most of the
CR interactions in the atmosphere cannot be simulated using these event
generators, we will see that the bulk of the produced mesons is still well
described due to the strong suppression of particle production near
threshold.

\begin{figure}[htbp]
	\centering
	\includegraphics[width=0.6\textwidth]{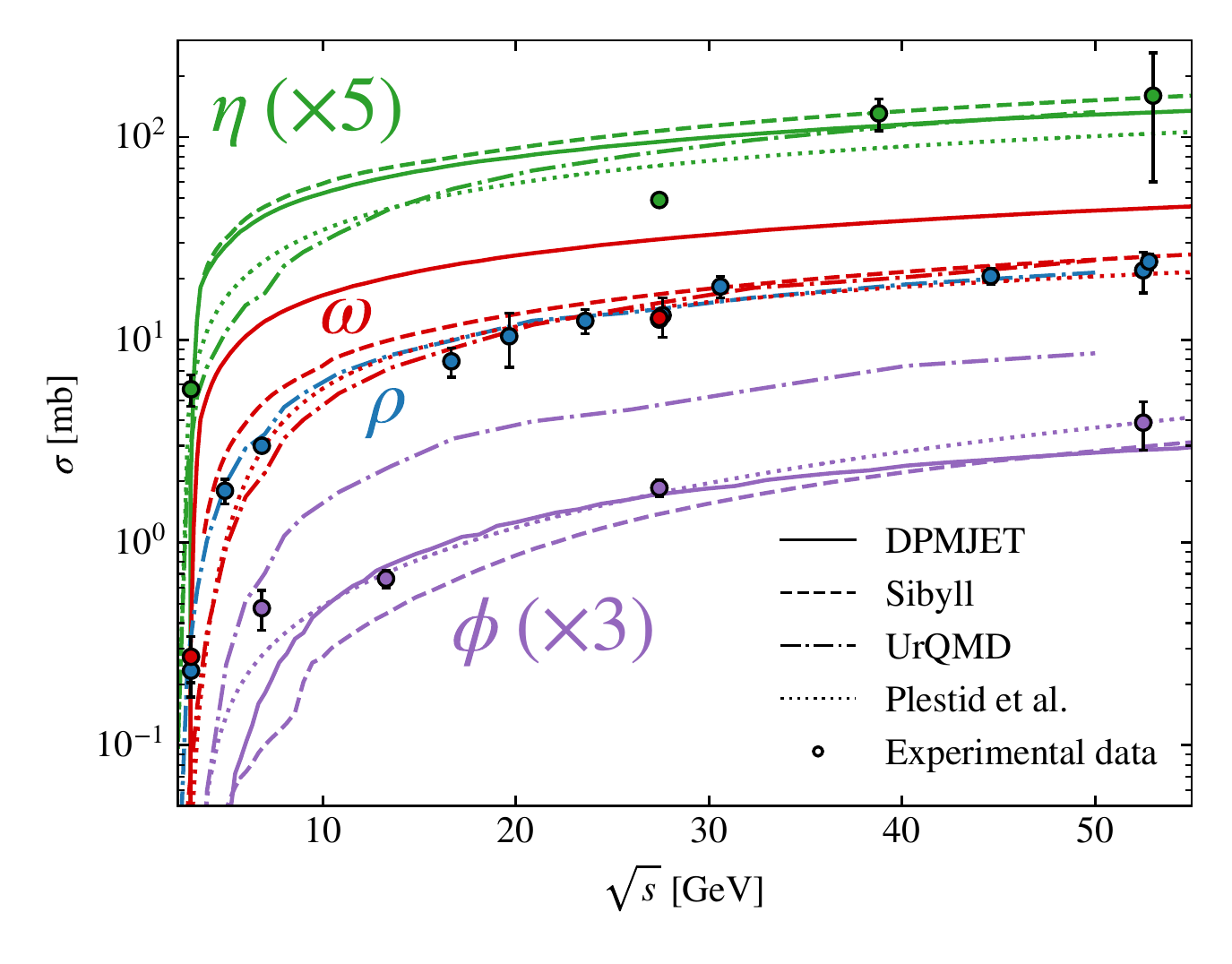}
	\caption{The total production cross section of $\eta$ (green),
          $\rho$ (blue), $\omega$ (red) and $\phi$ (purple) mesons computed
          using 
	DPMJET (solid), Sibyll (dashed) and UrQMD (dashed dotted)
	is compared to experimental data
	on the total inclusive cross sections from
    Refs.~\cite{HADES:2011ab,AguilarBenitez:1991yy,Baldini:1988ti}.
	The parametrisations used in Ref.~\cite{Plestid:2020kdm} are shown
	for comparison (dotted). The $\eta$ and $\phi$ cross sections are multiplied
	by a constant factor to make the figure clearer.
  In the case of Sibyll and DPMJET, the $\rho$ and $\omega$ fluxes
	are close to overlapping, thus only the $\omega$ flux is shown.
	}
	\label{fig:cross_section}
\end{figure}

To test the event generators, we compare in
Fig.~\ref{fig:cross_section} the production cross sections of 
$\eta$, $\rho$, $\omega$ and $\phi$  mesons computed using
DPMJET, Sibyll and UrQMD to the experimental data on the inclusive meson
production cross section $\sigma_{pp\to\meson X}$ in $pp$ collisions from
Refs.~\cite{HADES:2011ab,AguilarBenitez:1991yy,Baldini:1988ti}.
Additionally, we show the parametrisations used in
Ref.~\cite{Plestid:2020kdm}. There is overall a good
agreement between the experimental data and the predictions of the
event generators. We do, however, note a few deficiencies: First,
we see that UrQMD overproduces $\phi$ mesons by an energy-dependent
factor. Next, we note that DPMJET overproduces $\rho$ and $\omega$ mesons.
In this case, we obtain a good description of the data by
rescaling\footnote{%
  A proper solution which would require to increase appropriately
  the production cross sections of other particles as, e.g.,
  $a_0$ and $f_0$ mesons is planned for a future version of
  DPMJET~\cite{AF}.}
the production cross sections of $\rho$ and
$\omega$ mesons by a factor~0.5.
Finally, we comment on the case of
$J/\psi$ mesons: DPMJET predicts a $J/\psi$ production cross section that
is 3--4 orders of magnitude below those of Sibyll and Pythia, indicating that
the most important production channels of this meson are not included in this
event generator. 
We have therefore decided to focus in the following mainly on DPMJET and
Sibyll computing the production of $\eta$, $\rho$, $\omega$ and $\phi$
mesons, as they describe well the experimental data (after the rescaling)
and are reasonably fast. The
other aforementioned event generators will be used as basis for comparison.
In addition, Pythia will be used to describe the production of $J/\psi$
mesons.

\subsection{Flux of undecayed mesons}
\label{ssec:flux}

Next we compute the flux of undecayed mesons of the types
$\pi^0$, $\eta$, $\rho^0$, $\omega$, $\phi$ and $J/\psi$
in hadronic air showers induced by cosmic ray proton and helium nuclei
using a self-written Monte Carlo code in which
the interactions of hadrons are handled using QCD inspired
event generators. 
We consider He, $p$, $\pi^\pm$ and $K^\pm$ as stable\footnote{At low energies,
  charged mesons will decay, implying that this
treatment will lead to an overestimation of the meson yields. 
Considering for concreteness $E_c=30$~GeV \cite{Engel:2011zzb} as a 
hard cut-off and $\pi^\pm$ as primaries, we can estimate that there will
only be an effect at meson energies
$E<(m_N^2+m_{\pi^\pm}^2+2m_NE_c)^{1/2}-m_N-m_{\pi^\pm}\simeq 6.5$\,GeV.
The effect will be small as there will be a large contribution from
charged pions at higher energies. In particular, the effect on interesting
observables in the production of exotic particles above detector thresholds
will be negligible.}
projectiles and
nitrogen as target\footnote{%
	With Pythia, we consider only $pp$ interactions and take
	into account the helium flux by rescaling the
	proton flux appropriately.
}.
All short-lived particles that are not treated as a projectile are set to
decay using the decay subroutines in Pythia. 
Since we are mainly interested in the integrated meson fluxes
as a function of energy, we neglect the direction of the produced particles,
keeping all down-going particles. Moreover, 
practically all primaries above the production threshold will interact, and we
can therefore ignore the finite extension of the atmosphere.
Finally, we can neglect the tertiary contribution to meson production
due to the electromagnetic shower component, because
the average energy per produced secondary is for photons/electrons
smaller than for mesons and the cross sections for, e.g., photo-pion
production are suppressed relative to electromagnetic ones.
For the primary CR fluxes, we use the parametrisations fitted
to proton and helium data from AMS-02, DAMPE and CREAM
given in Ref.~\cite{Kachelriess:2020uoh}.

Before proceeding, the chosen energy cutoffs should be discussed.
We use $2\unit{GeV}/n$ as a low-energy threshold for DPMJET, QGSJET and
UrQMD, while we set $8\unit{GeV}/n$ and $60\unit{GeV}/n$ for Sibyll\footnote{
	The chosen threshold for Sibyll is lower that the intended validity
	range, but as seen in subsection~\ref{app:cross_section} Sibyll describes well the
	production cross sections down to $E_\mathrm{lab}\simeq 6.4\unit{GeV}$.
}
and Pythia, respectively. These energy cutoffs should be compared to
the threshold energy in the interaction $pp\to pp\meson$ which are 1.2, 2.2,
2.8, 3.5 and 12\,GeV for $\pi^0$, $\eta$, $\rho^0$, $\omega$, $\phi$ and
$J/\psi$ mesons, respectively.
Thus the chosen cut-off in DPMJET, QGSJET and UrQMD
is sufficiently small for all considered mesons except for the $\pi^0$;
the results for this meson must therefore be considered with
care\footnote{
	Note that we consider $\pi^0$ mesons only for completeness.
	As we will see in the next section, exotic particle production
	from $\pi^0$ decays is already strongly
	constrained by collider experiments. It is therefore doubtful that
  the decay of $\pi^0$ in atmospheric cascade can give leading
    constraints.
}.
Even more, the threshold suppression in the production cross section
will at some point be stronger than
the power law increase in the
primary CR flux. This means that even if Sibyll cannot describe most
particle interactions, it will still describe the bulk of produced mesons
more massive than $\pi^0$. Likewise, Pythia will describe well the
atmospheric $J/\psi$ flux, as is readily seen by Fig.~2 in
Ref. \cite{Plestid:2020kdm}.

\begin{figure}[htbp]
	\centering
	\includegraphics[width=0.6\textwidth]{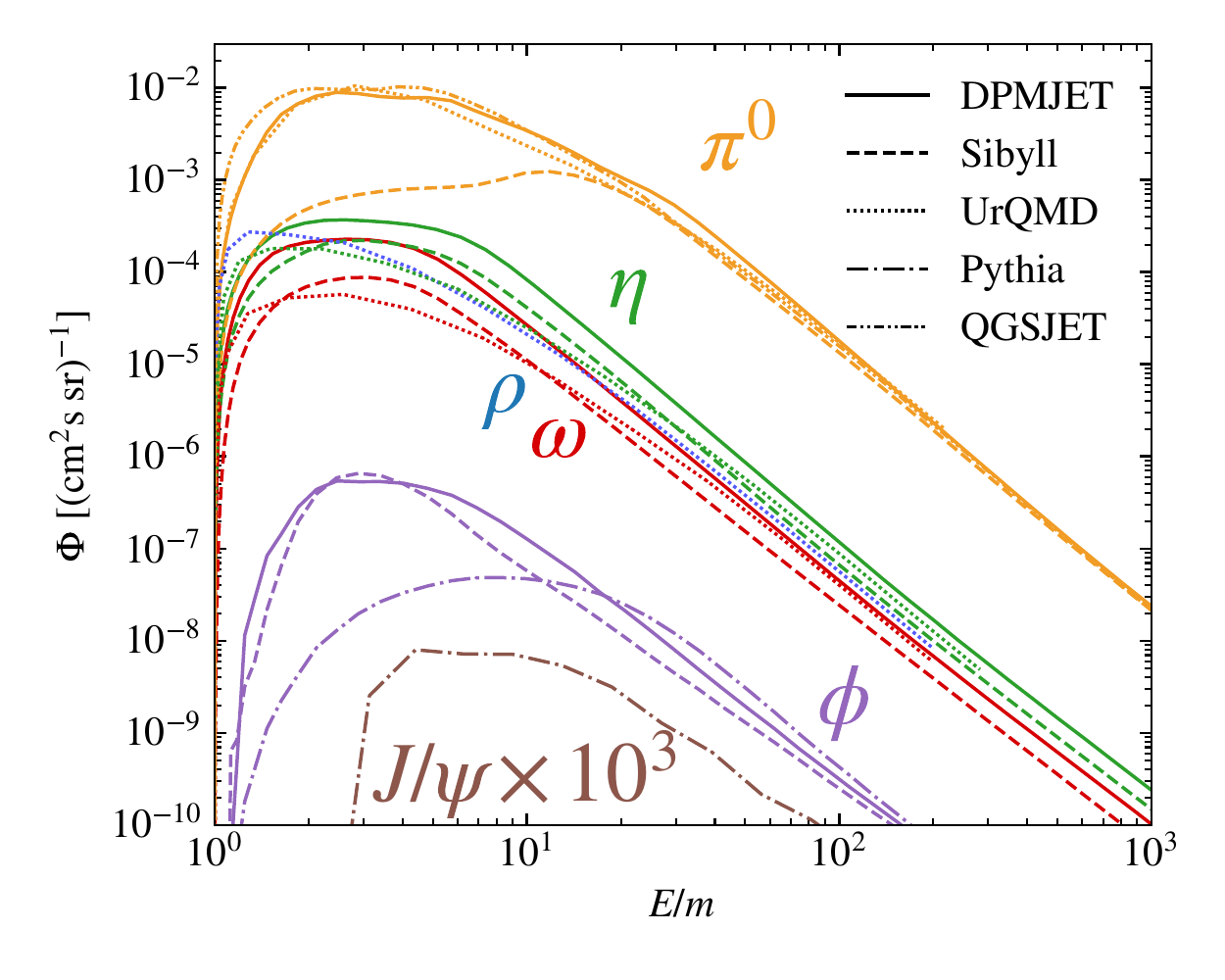}
	\caption{
	The flux of produced $\pi^0$ (orange), $\eta$ (green), $\rho$ (blue),
	$\omega$ (red), $\phi$ (purple) and $J/\psi$ (brown) mesons
	in $pp$ and He$p$ collisions at the top of the atmosphere
	are computed using DPMJET (solid lines), Sibyll (dashed),
	QGSJET (dashed-dotted-dotted), Pythia (dashed-dotted)
	and UrQMD (dotted). Only a selection of meson species are
	shown for different event generators to make the figure clearer.
	In the case of Sibyll and DPMJET, the $\rho$ and $\omega$ fluxes
	are close to overlapping, thus only the $\omega$ flux is shown.
	}
	\label{fig:simple}
\end{figure}

The main contribution to the meson production comes from the first
interaction at the top of the atmosphere, because the cosmic ray flux
is a steeply falling function of energy, $\Phi^\mathrm{CR}(E)\propto E^{-2.7}$.
Moreover, the cosmic ray flux is at low energies dominated by protons.
Therefore, we start by plotting in Fig.~\ref{fig:simple}
the integrated meson fluxes from $pp$ interactions weighted by the cosmic ray
flux $\Phi^\mathrm{CR}(E)\simeq\Phi_p(E) + 4\Phi_\mathrm{He}(E/4)$
as a simple benchmark case. 
Note that the production yields of $\rho$ and $\omega$ mesons are divided
by a factor 2 in the case of DPMJET, as described in the previous subsection.
As a basis for comparison, we plot also the result obtained using
QGSJET (only for $\pi^0$),
Pythia (only $\phi$ and $J/\psi$) and UrQMD (only $\pi^0$, $\eta$,
$\rho$ and $\omega$).
The effect of the chosen cutoffs are clearly visible: Lowering the
energy threshold extends the power law to lower energies and increases
the flux of produced mesons at low energies. For instance, the maximum of the
$\pi^0$ flux computed with Sibyll is suppressed by a factor $\simeq 10$
compared to DPMJET. This effect is smaller for heavier mesons, because of
their increased production threshold. The remaining overall differences for
heavier mesons can be explained by the differences
in the computed production cross sections (see section \ref{app:cross_section}).
The differences between DPMJET and Sibyll (for mesons heavier than $\pi^0$) 
capture well the uncertainties in the different event generators, which are
below a factor 2--3.

\begin{figure}[htbp]
	\centering
	\includegraphics[width=0.6\textwidth]{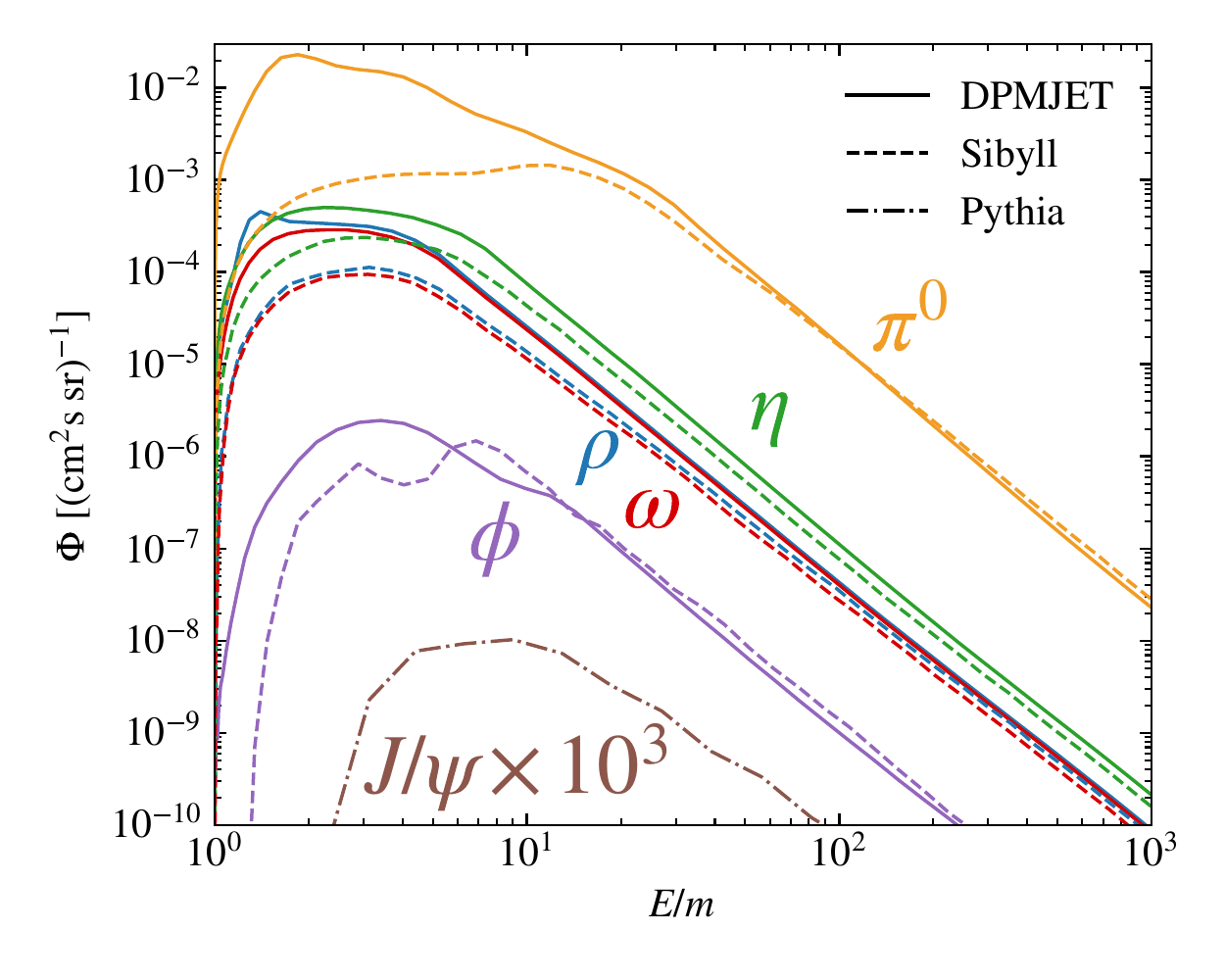}
	\caption{Meson flux produced in air showers using DPMJET, Sibyll and 
	Pythia (only $J/\psi$).
	The line-styles are the same as in Fig.~\ref{fig:simple}.
	}
	\label{fig:helium}
\end{figure}

\begin{figure}[htbp]
	\centering
	\includegraphics[width=0.6\textwidth]{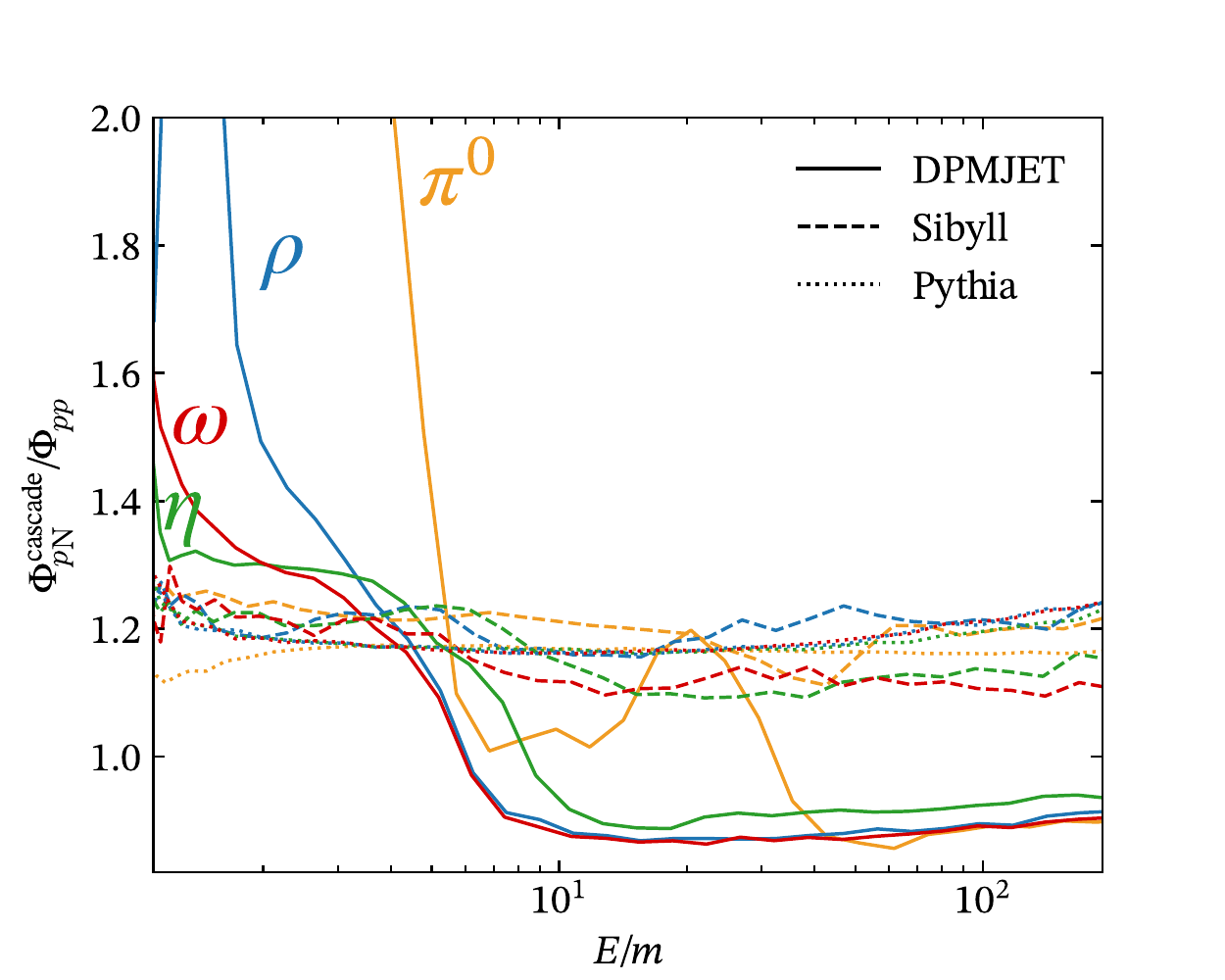}
	\caption{Ratio $\Phi^\mathrm{cascade}_{pN}/\Phi_{pp}$ of the meson
          fluxes in $pN$ air showers and a single $pp$ collision 
          computed using DPMJET (solid lines) and Sibyll (dashed lines).
          For comparison, $\Phi^\mathrm{cascade}_{pp}/\Phi_{pp}$ of the fluxes
          from a $pp$ air shower and a single $pp$ interaction  is shown in
          the case of Pythia. The color-scheme indicating the various mesons
          are the same as in the previous figures.
	}
	\label{fig:ratio}
\end{figure}

In Fig.~\ref{fig:helium}, the computed undecayed meson fluxes from hadronic
air showers are shown. The effect of including the cascade leads to a noticeable
increase in the flux at low meson energies, and shifts its maximum slightly to
smaller energies. However, the difference in the fluxes for Sibyll and DPMJET 
are larger than the gain in including the complete cascade, even at low
energies. These effects are more visible in Fig.~\ref{fig:ratio}. Here, 
the ratios of the meson fluxes of $\pi^0$, $\eta$, $\rho$ and $\omega$
from $p$N initiated air showers, $\Phi^\mathrm{cascade}_{pN}$,
and from a single $pp$ interaction at the top of
the atmosphere, $\Phi_{pp}$, are shown. For comparison, the ratio
$\Phi^\mathrm{cascade}_{pp}/\Phi_{pp}$ of the fluxes from a $pp$
air shower and a single $pp$ interaction is shown for Pythia.
Interestingly, the effect of including target nuclei
lowers the flux at large energies, because the kinetic energy
in the center of mass frame of the interaction is effectively reduced.
This effect is larger for DPMJET than for Sibyll.

The meson fluxes in Fig.~\ref{fig:helium} differ significantly\footnote{
	One should note that the meson fluxes found in Ref.~\cite{Plestid:2020kdm}
	were considered as \textit{a useful byproduct}, while their main
	result---the integrated flux of mCPs above detector thresholds---are
	mostly sensitive to the total production cross sections.
}
from those computed in Ref.~\cite{Plestid:2020kdm}:
For small meson energies, our fluxes
are suppressed more strongly, leading to a difference of about one order of
magnitude around the maximum. Meanwhile, for large meson energies the
differences are small and consistent with the differences in the production
cross sections discusses in subsection~\ref{app:cross_section}.

\section{Application: Atmospheric production of millicharged particles}
\label{sec:application}

In this section we analyze the production of mCP from the intermediate meson
decays in the atmosphere. This serves as a (conservative) benchmark model for
mCP, with the advantage of having only two free parameters: its mass $m$ and
charge $e\epsilon$. We take into account the decays
$\pi^0\to\{\bar{\chi}\chi, \bar{\chi}\chi\gamma\}$,
$\eta\to\{\bar{\chi}\chi, \bar{\chi}\chi\gamma\}$,
$\rho\to \bar{\chi}\chi$,
$\omega\to\{\bar{\chi}\chi, \bar{\chi}\chi\pi^0\}$,
$\phi\to\bar{\chi}\chi$ and
$J/\psi\to \bar{\chi}\chi$.
The corresponding branching ratios are estimated by rescaling the 
dilepton and diphoton branching ratios, as explained in \ref{app:BR}.
We handle the $1\to 3$ decays using the decay subroutines in Pythia 8, 
whereas the momentum distribution
in the $1\to 2$ decays is taken to be monoenergetic and isotropic in the rest
frame of the mother particle.

\begin{figure}[htbp]
	\centering
	\includegraphics[width=0.6\textwidth]{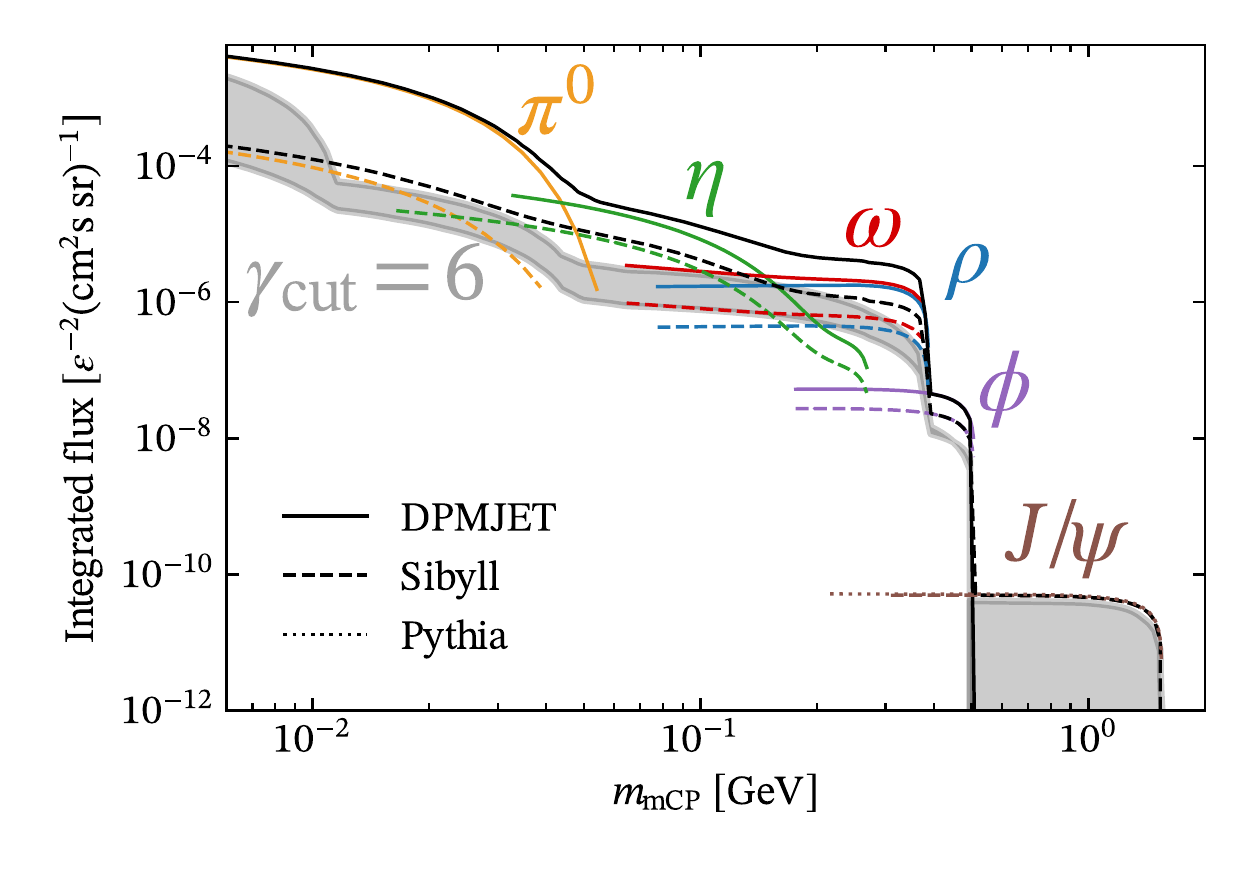}
	\caption{The integrated flux of
  mCPs from meson decays in the atmosphere
  for varying mCP mass using
	DPMJET (solid), Sibyll (dashed) and Pythia (dotted; only $J/\psi$)
	are shown in black.
	The contributions from the different meson species are
	indicated using the same colors as in Fig.~\ref{fig:simple}.
	The shaded gray region indicates the integrated flux above
	$\gamma_\mathrm{cut}=6$.
	}
	\label{fig:mCP}
\end{figure}

The integrated mCP fluxes computed using DPMJET and Sibyll are shown in
Fig.~\ref{fig:mCP} as a function of the mCP mass. The result obtained
for $J/\psi$ using Pythia is also shown. The step-like
behaviour arises from the various thresholds at $m_\meson/2$, as indicated
in the figure.
In addition, we show
the integrated mCP flux with a cutoff at the Lorentz factor
$\gamma_\mathrm{mCP}=E/m_\mathrm{mCP} = 6$, which 
corresponds approximately to the cut-off of the Super-Kamiokande experiment
used in their search for relic supernova neutrinos \cite{Bays:2011si}
(see~\ref{sec:upper_limit}): The upper line of the shaded gray region
corresponds to the mCP flux for $\gamma_\mathrm{mCP}>6$ calculated with DPMJET,
while the lower line uses Sibyll. 

\begin{figure}[htbp]
	\centering
	\includegraphics[width=0.6\textwidth]{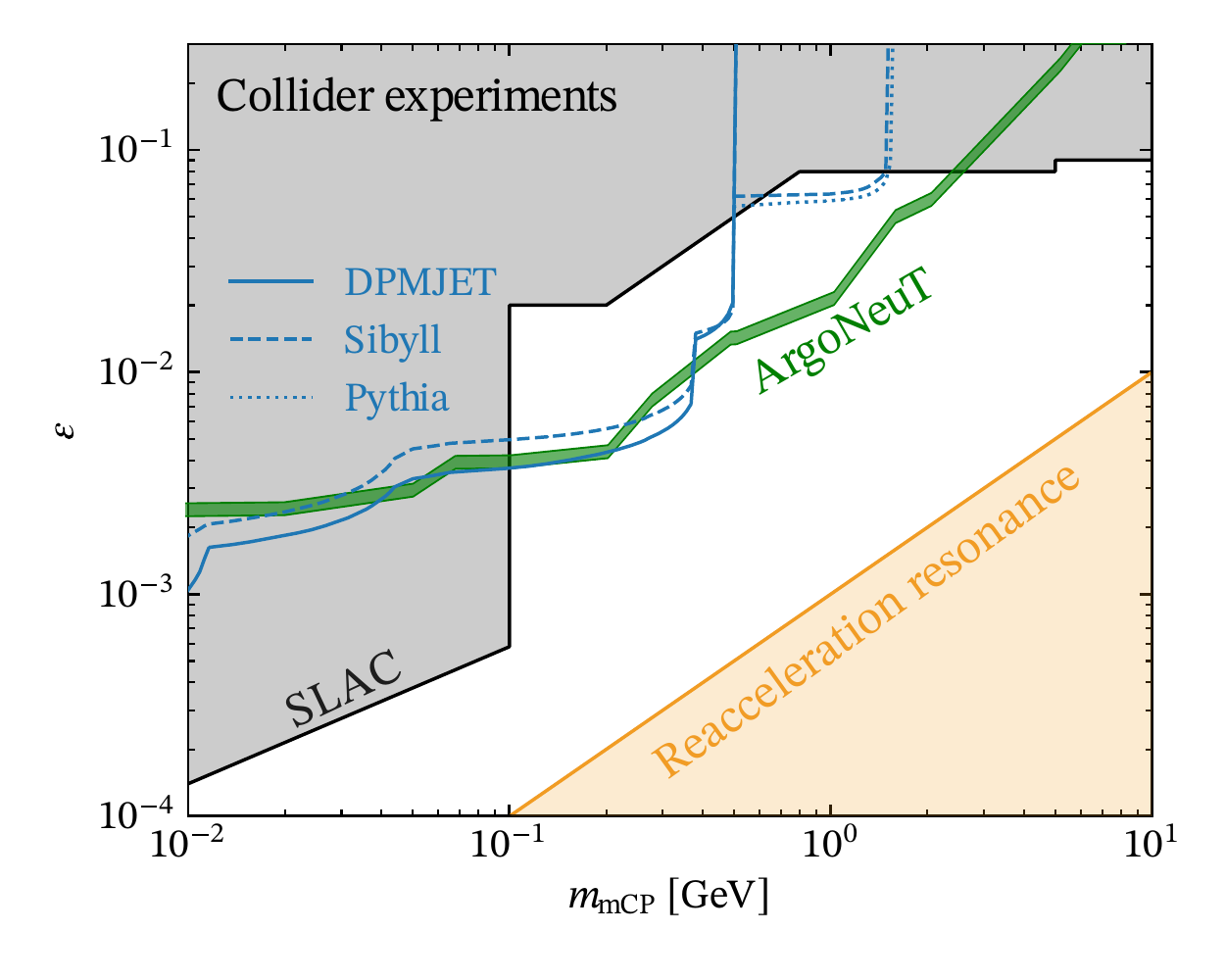}
	\caption{The upper limit in $(m, q)$-parameter space of mCPs
	set by the non-detection supernova neutrino events in
	Super-Kamiokande \cite{Bays:2011si}. The result is similar to
	that of Ref. \cite{Plestid:2020kdm}.
	To put it in perspective, the upper limit set by various collider
	experiments
	\cite{Prinz:1998ua,CMS:2012xi,Davidson:2000hf,Golowich:1986tj,Davidson:1991si}
	and the ArgoNeuT experiment \cite{Acciarri:2019jly}
	are shown. The strong limit below the $\pi^0$ threshold comes from the
	search for millicharged particles at SLAC \cite{Prinz:1998ua}.
	The lower limit set by the
	reacceleration condition \cite{Kachelriess:2020ams} is only valid if the
	mCPs make up more that $\sim 10^{-6}$ of the relic abundance.
	}
	\label{fig:exclusion}
\end{figure}

For completeness, we also show an exclusion plot using data from
Super-Kamiokande~\cite{Bays:2011si} employing the procedure introduced in
Ref.~\cite{Plestid:2020kdm}. A brief description of the procedure is given in 
\ref{sec:upper_limit}. The result is shown in Fig.~\ref{fig:exclusion} where it
is compared to a minor subset of existing limits (see e.g.\
Ref.~\cite{Magill:2018tbb} for additional bounds).
Intriguingly, the limit set by atmospheric
mesons is comparable to the existing strong limit set by the 
ArgoNeuT experiment~\cite{Acciarri:2019jly}. Note also that neutrino detectors
may have a significantly lower threshold energy than used here for
Super-Kamiokande. For example, the 
Borexino detector has in principle a threshold of $\sim 200\unit{keV}$ only 
limited by the natural presence of $^{14}$C~\cite{Alimonti:2008gc}.
Thus, the limit set by atmospheric mesons could in principle be significantly
improved.
This strengthens the importance of an accurate description of atmospheric mesons,
and motivates future work on using neutrino detectors to search for
exotic physics as introduced in Ref.~\cite{Plestid:2020kdm}.

\section{Summary}
\label{sec:summary}

In this work we have computed the flux of atmospheric mesons by simulating
hadronic air showers using the event generators
DPMJET~III~19.1, Pythia~8.303, QGSJET~II-04, Sibyll~2.3d and UrQMD~3.4.
The emphasis was put on Sibyll and DPMJET,
as they describe well the total production cross sections and are fast.
Moreover, the difference between these two  event generators may serve
as an estimate for the theoretical uncertainties of our flux predictions.
We have focused on the production of mesons with large electromagnetic decay
channels, $\pi^0$, $\eta$, $\rho$, $\omega$, $\phi$ and $J/\psi$.

This work was motivated by Ref.~\cite{Plestid:2020kdm}, where the meson
fluxes produced in $pp$ collisions at
the top of the atmosphere were computed
by fitting various parametrisations
to measurements of meson production cross
sections. The obtained fluxes were in turn used to set constraints on generic
BSM models with a (meta-)stable millicharged component.
Our results are in good agreement with those of Ref.~\cite{Plestid:2020kdm}.
The largest differences (up to orders of magnitude at low meson energies) arise
due to the different treatment of the cross sections.
The Monte Carlo approach has several advantages compared to using
parametrisations: secondary interactions as well as the effect of
interacting nuclei can be taken into account. Moreover,
no assumption has to be made
on the momentum dependence of the differential cross section.
By comparing the results obtained using DPMJET and Sibyll
for mesons more massive than $\pi^0$ we found
an estimated uncertainty of the theoretical predictions of a
factor 2--3. This uncertainty is larger than the changes resulting from
adding secondary interactions and simulating helium as projectile and
nitrogen as target. The effect of the larger interaction threshold in Sibyll
compared to DPMJET is small for all considered mesons, except for
$\pi^0$ at low meson energies.

Our tabulated results can be used to evaluate the flux of exotic particles
produced by atmospheric meson decays within generic extensions of the standard
model. As an example, we re-considered the production of millicharged
particles, and found that their production in atmospheric meson
decays can set leading constraints in the possible charge-to-mass ratio. 
The in principle lower thresholds for neutrino experiments is a strong
motivation for continuing the study of exotic particles produced in meson
decay in the atmosphere.

\section*{Acknowledgments}
\noindent
We would like to thank Anatoli Fedynitch for helpful comments on DPMJET~III.

\appendix

\section{Meson branching ratios into millicharged particles}
\label{app:BR}

The branching ratios of mesons $\meson$ into mCPs $\chi$ can be found by
rescaling the branching ratios of their electromagnetic decay channels into
charged leptons $l^\pm$.
The direct decay of mesons into mCPs, $\meson \to\bar{\chi}\chi$, the Dalitz
decays of pseudoscalar mesons, $P\to \bar{\chi}{\chi}\gamma$, and the
three-particle decay of vector mesons, $V\to \bar{\chi}{\chi}P$ will give the
dominant contributions to the mCP production in hadronic interactions. We use 
for evaluation\footnote{We use, except for $\pi^0$ decays, muons for
  comparisons, $l^\pm=\mu^\pm$.} of the relevant branching ratios the
experimental values given by the particle data group~\cite{Zyla:2020zbs}.

The branching ratio of a direct meson decay into two mCPs,
$\meson \to\bar{\chi}\chi$,
can be found by rescaling the dilepton branching ratio~\cite{Landsberg:1986fd}
as
\begin{equation}
	\frac{\br{\meson\to\bar{\chi}\chi}}{\br{\meson\to l^+l^-}}
	=\epsilon^2\sqrt{\frac{1-4m_\chi^2/m_\meson^2}
		{1-4m_l^2/m_\meson^2}}\frac{1+2m_\chi^2/m_\meson^2}
		{1+2m_l^2/m_\meson^2}.
	\label{eq:br2}
\end{equation}
In this work, we consider the direct decays of $\pi^0$, $\eta$, $\rho$,
$\omega$, $\phi$ and $J/\psi$.

The branching ratio of a pseudoscalar mesons into a photon and a mCP pair,
$P\to\gamma\bar{\chi}\chi$, can be computed by
rescaling the diphoton branching ratio~\cite{Landsberg:1986fd} as
\begin{equation}
\frac{\br{P\to\gamma\bar{\chi}\chi}}{\br{\meson\to\gamma\gamma}}=
	\frac{2\alpha\epsilon^2}{3\pi}\int_{4m_\chi^2}^{m_\meson^2}\dd{q^2}
	\sqrt{1 - \frac{4m_\chi^2}{q^2}}
	\left(1 + 2\frac{m_\chi^2}{q^2}\right)
	\frac{1}{q^2}
	\left(1-\frac{q^2}{m_\meson^2}\right)^3
	|F_\meson(q^2)|^2,
\end{equation}
with $F_\meson(q^2)$ being the meson form factor and $q^2$ the invariant
mass of the virtual photon. Likewise, the branching ratio for a vector
(pseudoscalar) meson into two mCPs and a pseudoscalar (vector) meson is
given by
\begin{equation}
\begin{aligned}
	\frac{\br{\meson\to A\bar{\chi}\chi}}{\br{\meson\to\gamma\gamma}} =
	\frac{\alpha\epsilon^2}{3\pi}\int_{4m_\chi^2}^{(m_\meson-m_\chi)^2}\dd{q^2}&
	\sqrt{1 - \frac{4m_\chi^2}{q^2}}
	\left(1 + 2\frac{m_\chi^2}{q^2}\right)
	\frac{1}{q^2}
	\\ &\times\left[\left(1 + \frac{q^2}{m_\meson^2 - m_A^2}\right)^2 -
	\frac{4m_\meson^2 q^2}{(m_\meson^2 - m_A^2)^2}\right]^{3/2}
	|F_\meson(q^2)|^2.
\end{aligned}
	\label{eq:brd2}
\end{equation}
In this work, we consider the three-body decays $\pi^0\to\gamma\bar{\chi}\chi$,
$\eta\to\gamma\bar{\chi}\chi$ and $\omega\to\pi^0\bar{\chi}\chi$.
We take into account the meson form factors using the parametrisations
from Refs.~\cite{Landsberg:1986fd,Zyla:2020zbs}:
\begin{equation}
	F_{\pi^0}(q^2)\approx 
	1 + q^2b_{\pi^0}, \quad b_{\pi^0}= (5.5\pm 1.6)\unit{GeV^{-2}},
\end{equation}
and
\begin{equation}
	\begin{aligned}
	F_i(q^2) = \left(1 - \frac{q^2}{\Lambda_i^2}\right)^{-1}, \quad
		&\Lambda_\eta = (0.716\pm 0.011)\unit{GeV},\\
		&\Lambda_\omega = 0.65\unit{GeV}.
	\end{aligned}
\end{equation}

\section{Upper limit from neutrino experiments}
\label{sec:upper_limit}

Water Cherenkov detectors like Super-Kamiokande~\cite{Bays:2011si} search
for the light signal emitted by the relativistic charged particles.
Therefore the light signal emitted by the scattered electrons in the
elastic interactions $\chi e^-\to \chi e^-$ can be used to constrain
the flux of mCP with scatterings
within a kinetic energy range $T_\mathrm{min}<T<T_\mathrm{max}$. This leads
to a ``windowed cross section'' for mCP-electron interactions
that can approximated as
\begin{equation}
	\tilde{\sigma}_{e\chi}(\gamma_\chi) =\int_{q_\mathrm{min}^2}^{q_\mathrm{max}^2}
	\dv{\sigma_{e\chi}}{q^2}\dd{q^2}\approx
	\frac{2\pi\alpha^2\epsilon^2}{2T_\mathrm{min}m_e}
	\left(1 - \frac{T_\mathrm{min}}{T_\mathrm{max}}\right)
	\Theta(\gamma_\chi-\gamma_\mathrm{cut})
\end{equation}
with $\gamma_\mathrm{cut}\approx 0.6\sqrt{2T_\mathrm{min}/m_e}+
0.4\sqrt{2T_\mathrm{max}/m_e}$ \cite{Plestid:2020kdm}. The expected number of
events is then
\begin{equation}
	N_{e\chi}\approx N_et\int_{\gamma_\mathrm{cut}}^\infty
	\tilde{\sigma}_{e\chi}(\gamma_\chi)\dv{\Phi_\chi}{\gamma_\chi}(\gamma_\chi)
	\dd{\gamma_\chi}
\end{equation}
with $N_e$ as the number of electrons in the detector and $t$ as the sampling 
period. For Super-Kamiokande, $T_\mathrm{min}=16\unit{MeV}$ and
$T_\mathrm{max}=88\unit{MeV}$ \cite{Bays:2011si},
corresponding to $\gamma_\mathrm{cut}\approx 6$ \cite{Plestid:2020kdm}.
Since the event shape of mCPs is similar to that of the supernova
background (see Fig. 10 in Ref. \cite{Plestid:2020kdm}), one can make use of
the analysis performed in Ref. \cite{Bays:2011si} for Super-Kamiokande,
which essentially leads to an exclusion of  $\sim 4$ events are excluded
with $90\unit{\%}$ CL.

\end{document}